\documentclass[usegraphicx,useAMS,usenatbib]{mn2e}
\title{The consequence of  jet interacting with a warped disc}
\author[Zuo and Gong]
{S.F. Zou$^{1}$ and B.P. Gong$^{1}$\thanks{E-mail: bpgong@mail.hust.edu.cn} \\
$^1$ Department of Physics, Huazhong University of Science and Technology, Wuhan 430074, China \\
}

\begin{document}

\date{Accepted, Received ;}

\pagerange{\pageref{firstpage}--\pageref{lastpage}} \pubyear{2002}

\maketitle

\label{firstpage}

\begin{abstract}
{The sprinkler pivots on a bearing on top of its threaded attachment nut. It is driven in a circular motion by a spring-loaded arm pushed back by the water stream which returns to ``impact" the stream. The water stream can thus rotate around a fix axis.
Analogously in our universe, the outflow or jet  formed by the relativistic plasma corresponds to the water stream of a sprinkler; and the baryons
in the tilted  outer accretion disc or torus play the role of ``impact arm".  Then the jet aligning with the inner parts of a warped disc can directly ``touch" the outer region of the disc.
The resultant collision between such rapid leptons and slow baryons automatically accounts for the main features of broad-line region of active galactic nuclei. Moreover, it naturally provides a channel  of dissipating the angular momentum of an accretion disc, which has long been a problem in theory of accretion disc.
}
\end{abstract}


\section{Introduction}

%%%%%%%%%%%%%%%%%%%%%%%%%%%%%%%%%%%%%%%%%%%%%%%%%%%%%%%%%%%%%%%%%%%%%%%%%%%%%%%%%%%%%%%%%%%%%%%%%%%%%%%%%%%%%%%%%%%%%%%%%%%%%
An  active galactic nucleus (AGN) is  comprised of three major elements, an accretion disc, a central spinning black hole (BH), and a jet.
The complex interactions among them result in fascinating phenomena of all bands, from radio, optical to X-ray and Gamma-ray, and displaying in all forms, from images to spectra and time varying effects.
Although tremendous progresses have been made in understanding the mechanisms underlying them,  many fundamental questions are still  open.

Firstly, frame dragging produced by a rotating BH with angular momentum $S$ causes precession of a particle if its orbital plane is inclined in relation to the equatorial plane of the rotation object. This is known as the Lense-Thirring effect \citep{LT}, with a   precession velocity, ${\Omega_{LT}\propto
1/r^{3}}$, of strong function of radius, $r$, so that the inner parts of the disc tends to align with the direction of the BH spin.

In the case where the BH spin is misaligned with the rotation direction of the disc, the disc becomes twisted and warped.
Such a warp can  propagate in the disc in a diffusive manner through viscosity  \citep{Pr}. And the warp  can make a large increase in the accretion rate \citep{LP}.

On the other hand, when  a BH  is in a binary system,   the inner regions of the disc remain aligned with the spin of the central BH. Whereas, the outer region precesses around the  orbital axis, owing to the tidal torque of the companion object   \citep{Ka,Sill,Ka1,Rome,CA,MPT}. This results in  a warped and twisted disc.

The apparent lack of correlation between the orientation of the radio jets and the plane of the host galaxies disc might also be attributed to warped discs  \citep{Sch}.

Then there is a question: what if a disc is so warped that the jet axis aligning with the inner disc touches the outer disc?
Once  such a warp induced jet-disc interaction occurs, the inner and outer parts of a warped disc interact in a new way other than the viscosity,
which influences  the precession velocity of jet,  accretion rate and stability of the disc.

Secondly, accretion disc has been proven to be one effective mechanism of extracting gravitational potential energy and converting it to radiation powering AGNs.
No matter a disc is warped or not, the accreted material at the inner region of the disc  has to dissipate its angular momentum in order to fall on to the BH.
If there is no external torque, the angular momentum loss is achieved through the accretion disc by outwardly  transferring  angular momentum.
Accretion onto an AGN black hole will be limited largely by how rapidly angular momentum can be  lost, which determines the power of AGN directly.
The angular momentum transferred to the periphery must have a reasonable way of loss, or the accumulation of external angular momentum would cause instability in the accretion disc.

Thirdly, one of the fundamental and defining characteristics of AGNs is the presence of broad emission lines in their optical and ultraviolet (UV) spectra. The investigation of the profile and intensities of these lines show that they originate in dense gas moving very fast around  the central engine, which are thought to be fully exposed to the intense radiation from the engine.

The outer region of disc and wind from disc as possible origins of BLR have been studied extensively  \citep{ELF}. However, the complexity of observations, e.g., the double-peaked broad emission lines are thought to originate in disc, while the local viscosity heating is usually insufficient for the observed luminosity of the emission site; and the correlation of BLR (optical and UV) with  Fe K$\alpha$ (X-ray)  lines  indicate that the origin of BLR is still an open question.

Interestingly, each of the three questions, the precession property, transfer of angular momentum, and origin of BLR, appears to be unrelated and difficult, but all of them together point to a simple scenario: sprinkler,  with the water stream replaced by the jet and ``impact arm" replaced by  the outer parts of  accretion disc.

The collision between jet (formed by relativistic electrons) and the outer disc (slow baryons), automatically kicked out baryons from the disc, which   provides a channel of angular momentum dissipation of a disc. Moreover, such a jet disc  interaction influences the overall precession velocity of the disc. Finally, the baryons in the disc, carrying velocity gained through the collision with electrons,  spread around the collision site, which
naturally account for the major features  of BLR.

Therefore,  sprinklers in our yards, may provide a clue  to understand what happens to ``sprinklers" of the garden of  universe.

%%%%%%%%%%%%%%%%%%%%%%%%%%%%%%%%%%%%%%%%%%%%%%%%%%%%%%%%%%%%%%%%%%%%%%%%%%%%%%%%%%%%%%%%%%%%%%%%%%555

\section{The New Model}

%% In a manner similar to \objectname authors can provide links to dataset
%% hosted at participating data centers via the \dataset{} command.  The
%% second curly bracket argument is printed in the text while the first
%% parentheses argument serves as the valid data set identifier.  Large
%% lists of data set are best provided in a table (see Table 3 for an example).
%% Valid data set identifiers should be obtained from the data center that
%% is currently hosting the data.
%%
%% Note that AASTeX interprets everything between the curly braces in the
%% macro as regular text, so any special characters, e.g. "#" or "_," must be
%% preceded by a backslash. Otherwise, you will get a LaTeX error when you
%% compile your manuscript.  Special characters do not
%% need to be escaped in the optional, square-bracket argument.

Once a disc is warped as shown in  Fig.$1$, the jet-disc interaction occurs. The collision  sites separate from  the central engine by a distance ${r_0}$.   The opening angle of the jet cone is denoted, ${\theta_0}$, and  the misalignment angle between the jet axis and the axis of the  precession cone is ${\alpha_0}$, as shown in  Fig.$1$.

%% In this section, we use  the \subsection command to set off
%% a subsection.  \footnote is used to insert a footnote to the text.


%% This section also includes several of the displayed math environments
%% mentioned in the Author Guide.

\section{Transfer of Angular Momentum}

At different radii of a disc, gaseous material is orbiting with differing angular velocities, $\Omega$, determined by Kepler's third law.
Because of the ever present chaotic thermal motions of fluid molecules or turbulent motions of fluid elements, viscous stresses are generated. This type of transport process is known as shear viscosity.

The torque exerted by the outer ring on the inner one (= $ - $the torque of the inner on the outer) is $G(R) = 2\pi R\nu \Sigma \mathop R\nolimits^2 \frac{{d\Omega (R)}}{{dR}}$, where $\nu$ is the kinematic viscosity, $\Sigma $ is the surface density of the disc, and $\Omega (R)$ is the angular velocity.
Such a  torque corresponds to a transfer of  angular momentum (per unit time) to the outermost ring,
\begin{equation}\label{Ldot}
\dot L \sim  - G(\mathop r\nolimits_0 ) =  - 2\pi \nu \Sigma \mathop R\nolimits^3 \frac{{d\Omega }}{{dR}}\mathop |\nolimits_{R = \mathop r\nolimits_0 }.
\end{equation}
The accretion rate of mass  $\dot M$ can be expressed by,
\begin{equation}\label{vS}
\nu \Sigma  = \frac{{\dot M}}{{3\pi }}[1 - (\frac{{\mathop R\nolimits_* }}{R}\mathop )\nolimits^{1/2} ],
\end{equation}
where $\mathop R\nolimits_*  = 2G\mathop M\nolimits_p /\mathop c\nolimits^2 $ is the Schwarzschild radius of a BH.

Substitute the Kepler angular velocity, $\Omega  = {\Omega _K}(R) = {(G{\rm{ }}{M_p}/{R^3})^{1/2}}$, and Eq.$\ref{vS}$ into Eq.$\ref{Ldot}$, the angular momentum loss rate can be written in terms of $\dot{M}$,
\begin{equation}\label{Ldot1}
\begin{array}{l}
\dot L \sim \{  - 2\pi \mathop R\nolimits^3  \cdot \frac{{\dot M}}{{3\pi }}[1 - (\frac{{\mathop R\nolimits_* }}{R}\mathop )\nolimits^{1/2} ]\mathop { \cdot \frac{{ - 3}}{2}(\frac{{G\mathop M\nolimits_p }}{{\mathop R\nolimits^5 }}\mathop )\nolimits^{1/2} \} |}\nolimits_{R = \mathop r\nolimits_0 } \\
{\kern 1pt} {\kern 1pt} {\kern 1pt} {\kern 1pt} {\kern 1pt} {\kern 1pt} {\kern 1pt} {\kern 1pt} {\kern 1pt} {\kern 1pt}  = (G\mathop M\nolimits_p \mathop )\nolimits^{1/2} \dot M(\mathop r\nolimits_0^{1/2}  - \mathop R\nolimits_*^{1/2} )
\end{array}.
\end{equation}
Such a $\dot{L}$ is the angular momentum  required to be transferred to outer disc per unit time in order to maintain persistence power of an AGN fueled by $\dot{M}$.
There must be an efficient channel  of angular momentum dissipation, otherwise the piled up angular momentum in the outer disc  can cause instability of the disc and stops the accretion process.

The sprinkler scenario provides a simple answer. The collision between jet and disc results in scattering  matters off the disc, which should be in Kepler motion around BH and hence carrying angular momentum.  Such a mass loss rate of the outer disc can be estimated by, $\dot{m}_{loss}=M_{disc}\dot{L}/L_{disc}$, where $M_{disc}=nm_p$ and $L_{disc}=nL_p$ are mass and angular momentum of the disc respectively ($n$ is the total number of baryons in the disc).

If we use the angular momentum of the outermost ring to represent that of the whole disc, the mass loss rate reduces to
\begin{equation}\label{mdot}
\dot{m}_{loss}\sim\frac{{{m_p}\dot L}}{{{L_p}}} = \dot M[1 - {(\frac{{{R_*}}}{{{r_0}}})^{1/2}}].
\end{equation}
where
$\mathop L\nolimits_p  = \mathop m\nolimits_p \mathop r\nolimits_0^2 \mathop \omega \nolimits_0  = \mathop m\nolimits_p (G\mathop M\nolimits_p \mathop r\nolimits_0 \mathop )\nolimits^{1/2} $ is the angular momentum of a proton. Thus,  a channel of dissipating angular momentum is set up, at  a warped disc  as shown in Fig.$1$.

\section{The Collision Process}

The jet-disc interaction dissipating the angular momentum can be calculated simply by the collision of relativistic leptons in the jet with the slow baryons in the outer disc, as shown in Fig.$2$.
Major parameters in the study of such a  jet-disc collision are: the volume of the interaction zone, $\mathop V\nolimits_i $; the number density of baryons (dominantly protons) in the outermost disc, $\mathop n\nolimits_p $; the number density of leptons (relativistic electrons) in the jet, $\mathop n\nolimits_e $, which travel at speed, $\mathop v\nolimits_e $. The number of leptons scattered into a solid angle between $\Theta $ and $\Theta  + d\Theta $ per unit time is given,
\begin{equation}\label{dNi}
d\mathop N\nolimits_i  = 2\pi \sigma (\Theta )\sin \Theta d\Theta  \cdot \mathop n\nolimits_e \mathop v\nolimits_e  \cdot \mathop n\nolimits_p \mathop V\nolimits_i .
\end{equation}
The number of leptons ejected out from the jet per unit time is estimated, $\mathop N\nolimits_e\sim {\xi}\dot M/\mathop m\nolimits_p$, where $\xi$ is a dimensionless coefficient, $\xi=1$ means accreted matter is totally ejected out.
With an opening angle of  jet, $\mathop \theta \nolimits_0 $, as shown in Fig.$1$, we have
${\rm{ }}{n_e}{\rm{ }}{v_e} = {N_e}/[\pi {(\frac{1}{2}{\rm{ }}{r_0}{\theta _0})^2}]$.
The change of linear momentum (per unit time) of   leptons involving in collision  is given, $d\mathop p\nolimits_e  = 2\sqrt {2\mathop m\nolimits_e E} \sin \frac{\Theta }{2}d\mathop N\nolimits_i$,  where $E = \mathop \gamma \nolimits_e \mathop m\nolimits_e \mathop c\nolimits^2 $ is the energy of a relativistic electron, with mass $m_e$ and Lorentz factor $\gamma_e$.
For simplicity, the Rutherford scattering cross section  \citep{GPS}, $\sigma (\Theta ) = \frac{1}{4}(\frac{{ZZ'\mathop e\nolimits^2 }}{{2E}}\mathop )\nolimits^2 \mathop {\csc }\nolimits^4 \frac{\Theta }{2}$, is used, where $E$ is the energy of the scattering particle.  Although it is derived in the case of  a repulsive scattering process, it is still a good approximation in the collision of the jet and disc.

Since the collision occurs predominantly between electrons and protons, the interacting charges satisfy, $(ZZ'\mathop )\nolimits^2  = 1$. Substituting Eq.$\ref{dNi}$ into the linear momentum change yields,
\begin{equation}\label{dpe1}
d\mathop p\nolimits_e  = \sqrt {2\mathop m\nolimits_e E} (\frac{{\mathop e\nolimits^2 }}{{2E}}\mathop )\nolimits^2 \frac{{8{\xi}\dot M\mathop n\nolimits_p \mathop V\nolimits_i }}{{\mathop m\nolimits_p \mathop r\nolimits_0^2 \mathop \theta \nolimits_0^2 }}\frac{{\cos \frac{\Theta }{2}}}{{\mathop {\sin }\nolimits^2 \frac{\Theta }{2}}}d\Theta ,
\end{equation}

The mean force corresponding to the interaction is
\begin{equation}\label{F}
F = \int_{\mathop \Theta \nolimits_0 }^\pi  {d\mathop p\nolimits_e }.
\end{equation}
The number density of protons in the outermost ring is $\mathop n\nolimits_p $, each proton occupies a volume $1/\mathop n\nolimits_p $, the functional relationship between the impact parameter and the scattering angle is  \citep{GPS}, $s = ZZ'{\rm{ }}{e^2}\cot \frac{\Theta }{2}/(2E)$. In such case, we have $s \le \mathop n\nolimits_p^{ - 1/3} $, the lower limit of the integral can be obtained by,
\begin{equation}\label{theta0}
\mathop n\nolimits_p^{ - 1/3}  = \frac{{ZZ'\mathop e\nolimits^2 }}{{2E}}\cot \frac{{\mathop \Theta \nolimits_0 }}{2}.
\end{equation}

Integrating Eq.$\ref{F}$, we get
\begin{equation}\label{F1}
F = [\sqrt {1 + (\frac{{2E}}{{\mathop e\nolimits^2 \mathop n\nolimits_p^{1/3} }}\mathop )\nolimits^2 }  - 1]\sqrt {2\mathop m\nolimits_e E} (\frac{{\mathop e\nolimits^2 }}{{2E}}\mathop )\nolimits^2 \frac{{16{\xi}\dot M\mathop n\nolimits_p \mathop V\nolimits_i }}{{\mathop m\nolimits_p \mathop r\nolimits_0^2 \mathop \theta \nolimits_0^2 }},
\end{equation}

Eq.$\ref{F1}$ corresponds to a torque, $N = F\mathop r\nolimits_0$, which makes the jet to precess around the core at angular velocity, $\vec{\Omega}_p$, determined by, $\vec{N}=\vec{\Omega}_p\times\vec{S}$, so that,
\begin{equation}\label{L}
 N=\Omega_p S \sin\alpha_0= \frac{{2\pi (1 + z) S \sin\alpha_0}}{{P}} ,
 \end{equation}
where $P$ is the precession period measured in the observer's reference frame, and $S = G{\rm{ }}M_p^2|{\rm{ }}{a_*}|/c$ is the spin angular momentum of the primary BH, with $\mathop a\nolimits_* $ its dimensionless angular momentum. Such a precession resembles  a sprinkler, with water stream and  impact arm replaced by the jet and the baryons at the outer disc respectively.

Thus, according to the collision process described by Eq.$\ref{dNi}$ to Eq.$\ref{L}$, the mass loss rate, $\dot{M}$, required to precess the jet at the period, $P$, is,
\begin{equation}\label{Vi}
\dot M = \frac{{\pi (1 + z){m_p}{r_0}\theta _0^2GM_p^2|{a_*}|\sin {\alpha _0}}}{{8c{n_p}{V_i}P\sqrt {2{m_e}E} {\rho ^2}[\sqrt {1 + {{(\rho n_p^{1/3})}^{ - 2}}}  - 1]}},
\end{equation}
where $\rho  = {e^2}/(2E)$. Fore simplicity, $\xi=1$ is assumed  in the calculation of Eq.$\ref{Vi}$. Notice that ${\dot M}$ of  Eq.$\ref{Vi}$ represents mass loss rate of the relativistic jet, which is thought to be comparable to the accretion rate of Eq.$\ref{Ldot1}$ derived from the angular momentum transfer of a disc.

The parameter spaces of ${\dot M}$, ${r_0}$ and $V_i$ corresponding to the two mechanisms given by Eq.$\ref{Ldot1}$ and Eq.$\ref{Vi}$, can be
displayed through the observational parameters of the Seyfert 1 galaxy 3C 120 and the quasar 3C 345, as shown in Table $\ref{Table:Param}$.

Assuming $\mathop \theta \nolimits_0  = \mathop {5}\nolimits^ \circ  $, ${\alpha _0} = {60^ \circ }$, $\mathop \gamma \nolimits_e  = 10$, ${\dot M_0} = 1\mathop M\nolimits_ \odot  /yr$, $\mathop n\nolimits_p  \sim \mathop {10}\nolimits^{30} \mathop m\nolimits^{ - 3} $, $|\mathop a\nolimits_* | = 0.9$, the cross area can be obtained  as shown in Fig.$3$.  Notice that in Fig.$3$ the inclined lines correspond to  Eq.$\ref{Vi}$; and the solid vertical ones correspond to Eq.$\ref{Ldot1}$ in which ${\dot L}$ is estimated under the binary model, with radius ${r_0}$ equivalent to the outer disc radius  ${r_d}$ under the tidal torque  \citep{PT,La}.

The shadowed area of Fig.$3$ corresponds to an acceptable parameters of collision process, the predicted torque of which can precess  the jet at a period  of tens of years, which is expected by Newtonian torque   \citep{Ri}.

The volume of the collision region is estimated,  ${V_i} \sim 2{r_0}{\theta _0} \cdot \pi R_{ring}^2$, the magnitude of which can be inferred from the  cross of vertical lines (correspond to different scenarios) with the inclined lines  of  Fig.$3$.

The shadowed area corresponds to the case that the 10yr precession period is driven by the Newtonian torque, so that the size of the outermost ring of the disc can be  estimated, ${R_{ring}} \sim {[{V_i}/(2\pi {r_0}{\theta _0})]^{1/2}}$, which is  $ \sim 3.5 \times {10^{11}}m$ for 3C 120, and $ \sim 7.4 \times {10^{13}}m$ for 3C 345.

In contrast, if the  $\sim$10yr precession period originates in the Lense-Thirring precession in a single BH, the jet-disc collision results in a twisted  outer ring of the disc. Because the collision makes the liner momentum of baryons in the outer disc vary as  Fig.$2$. Such a twisting of outer ring apparently changes with the jet precession, so that the whole disc, inner and outer, varies at the precession rate dominated by the inner disc.

\section{BLR Originated in Collision}

As a result of the jet-disc  collision,  baryons in the outer disc  scattered by high-speed leptons in jet spread around the collision site.

The momentum and energy exchange between slow baryons and  relativistic electrons are shown in Fig.$2$.
The velocity distribution of scattered baryons  can be calculated simply by   the  conservation of momentum,
\begin{equation}\label{v_cos}
v = 2\sqrt {2\mathop \gamma \nolimits_e } \frac{{\mathop m\nolimits_e }}{{\mathop m\nolimits_p }}c\cos \theta ,
\end{equation}
where ${\gamma _e}$ is the Lorenz factor of  relativistic electrons, and $\theta  = (\pi  - \Theta )/2$, as shown in  Fig.$2$, is the misalignment angle between  the velocity of protons and  the jet axis.

By Eq.$\ref{v_cos}$, the velocity gained by a baryon at one collision, is $\sim 10^3$km$/$s, which is responsible for the velocity inferred from the Doppler broadening of emission lines observed in AGNs.

Moreover, the jet-disc collision predicts a velocity distribution as shown in Fig.$4$, in  which magnitude of velocity decreases with the increase of $\theta$. Such a velocity distribution of baryons predicts   a bi-polar configuration of line emission.

In fact, Eq.$\ref{v_cos}$ is obtained ignoring the energy absorption by baryons, which can ionize the baryons. For simplicity, we can  assume that a half of energy before collision is absorbed by the baryons,  the magnitude  of the baryon velocity $v$ of Eq.$\ref{v_cos}$, will be  smaller. While  in  order of magnitude the conclusion drawn is unaffected.

Notice that the bi-polar corresponds to half opening angle of $(\pi  - {\Theta _0})/2$, and the smallest deflection angle ${\Theta _0}$ is given by Eq.$\ref{theta0}$.

In fact, Eq.$\ref{v_cos}$ predicts a correlation:  the velocity of baryons inferred from line  broadening  of BLR  is inversely related to precession period of the jet.

This is easy to understand. The more the linear momentum exchange between electrons of jet and the baryons of the disc, the more the momentum and energy gain for the baryons, so that they travel faster. On the other hand, this corresponds to larger action and reaction forces between jet and disc, and hence  larger torque on the jet, so that  the jet precesses more rapidly (and shorter precession period).
The rate of jet precession can be  approximated,  $\Theta  \sim 2\pi /P$, which is so small  that $v$ of Eq.$\ref{v_cos}$ can be written,
$v \approx \sqrt {2{\gamma _e}} \frac{{{\rm{ }}{m_e}}}{{{\rm{ }}{m_p}}}c\Theta  \propto \gamma _e^{1/2}/P$.
The predicted correlation is supported  by some AGNs with corresponding  observations, as  shown in Table $\ref{Table:short-period}$.

\section{Discussion}

Besides interpreting the difficulties in angular momentum transfer and precession velocity of jet, a number of observations concerning BLR likely favor the new model.

Firstly, the profiles of the broad lines observed at a high spectral resolution and signal-to-noise ratio, appear to be very smooth  \citep{ELF}.
The velocity distribution predicted by our model  originates in  the jet-disc collision, as shown in Fig.$4$, is automatically smooth.

Secondly, as shown in Fig.$1$, the new model actually implies a limited jet-disc interaction, in which only a small fraction of leptons in the jet is involved in the collision. Once the jet plunges  into the outer ring heavily, or touches part of inner rings, the collimated jet can be destroyed. This may explain the observations of radio quiet quasars, with the largest emission line width among AGNs, but  weak in radio emission.

Obviously, when a torus is outside the bi-polar configuration of BLR of an AGN , then obscurity of the BLR becomes more complicated. Therefore,
a LOS with a relatively small misalignment angle with the jet axis tends to observe single peaked emission line of BLR, as some Seyfert type 1.

And if  the  misalignment between LOS and jet axis is relatively large, the obscurity depends on the angle of LOS with the normal direction of the outer ring. As shown in Fig.$1$, a large misalignment  is favorable to observe double-peaked BLR, and a small one tends to be obscured, as  Seyfert type 2 (which is absence of  BLR emission).

Thirdly, a number of authors  \citep{CH,EH,Stra,Stra1} have pointed out that local viscous dissipation in the line-emitting region of accretion disc does not provide enough power to account for the observed emission line luminosity.

Apparently in our model this difficulty is naturally avoided.
During the collision, as shown in  Fig.$1$, baryons at the collision sites (or near)   gain energy and momentum (linear), which can automatically  ionize neutral hydrogen or other atoms.
Typically, a non-relativistic proton can  absorb  $\sim 1$eV in a head on collision with an electron of Lorentz factor 10 in the jet, which can well account for the energy required for line emission of BLR.
And multi-collision processes can easily excite a   hydrogen to UV band.

Moreover, if the jet contains certain fraction of  baryons with a bulk Lorentz factor of 10 also, their collisions with the baryons in the outer disc correspond to an energy exchange of $\sim 1$KeV, which is responsible for the energy required to invoke the line emission of  Fe K$\alpha$.

Because NGC 7213 displays  very close speeds  inferred from its  BLR and  Fe K$\alpha$ lines, it is used as an evidence of  BLR origin for  Fe K$\alpha$ line  \citep{BLM}.

From the point of view of the jet-disc interaction, the cross section of the relativistic electrons in the jet with the non-relativistic Hydrogen in the disc differs from that of  relativistic protons in jet and  non-relativistic  iron in the disc.

Therefore, even if a source shows  considerably discrepancy between   speeds  inferred from   BLR  and  Fe K$\alpha$ lines, the emission lines may still be originated at the same site of collision. So  evidence of common origin  should be found in  luminosity, number density, absorbtion extracted from the two types of emission lines.

Consequently, all these issues are realized at the expense of a warped disc, which makes the jet-disc interaction possible.




\section{Acknowledgments}
 We thank Y.C. Zou, Q.W. Wu, and D.X Wang  for useful discussion and comment. We also thank  J.Y. Gong for his help in Figure. This research is supported by the
National Natural Science Foundation of China, under grand
NSFC11178011.

\clearpage

\begin{figure}
\includegraphics[width=0.45\textwidth]{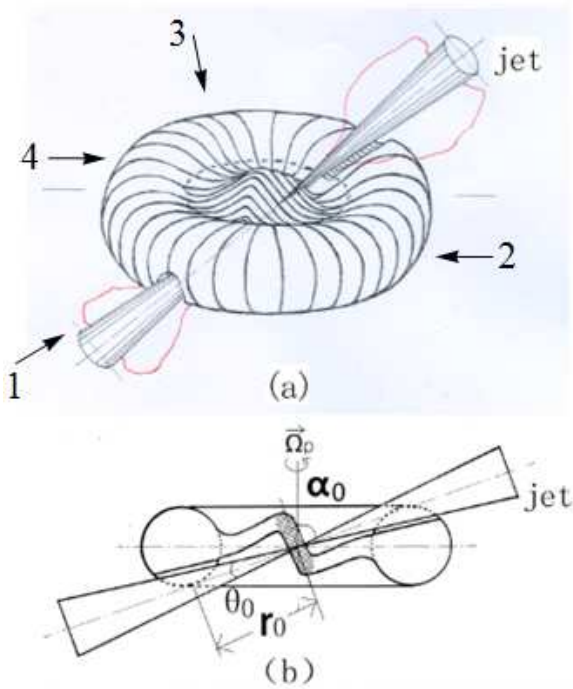}
%\epsscale{.50}
%\plotone{Fig.1.eps}
\caption{The interaction of a jet with a warped disc. At the collision site, baryons  in  the outermost ring of the warped disc are scattered off, which carry velocity of $\sim 10^3km/s$, and are responsible for line emission, as represented by the cloudy configurations.
In the figure, $\mathop r\nolimits_0 $ is the radius of the outermost ring, $\mathop \theta \nolimits_0 $ denotes the opening angle of the jet, $\mathop \alpha \nolimits_0 $ is the angle of the jet with the axis of the precession cone, and $\mathop \Omega \nolimits_p $ is the precession angular velocity. Observing at different view angles, the BLR line profiles appear different, e.g., 1 is likely responsible for the single-peaked one; 3 for double-peak; and  2, 4 for absence of BLR lines (obscured by the warped disc or an outer torus co-plane with the outer disc).\label{fig1}}
\end{figure}

\clearpage

\begin{figure}
\includegraphics[width=0.45\textwidth]{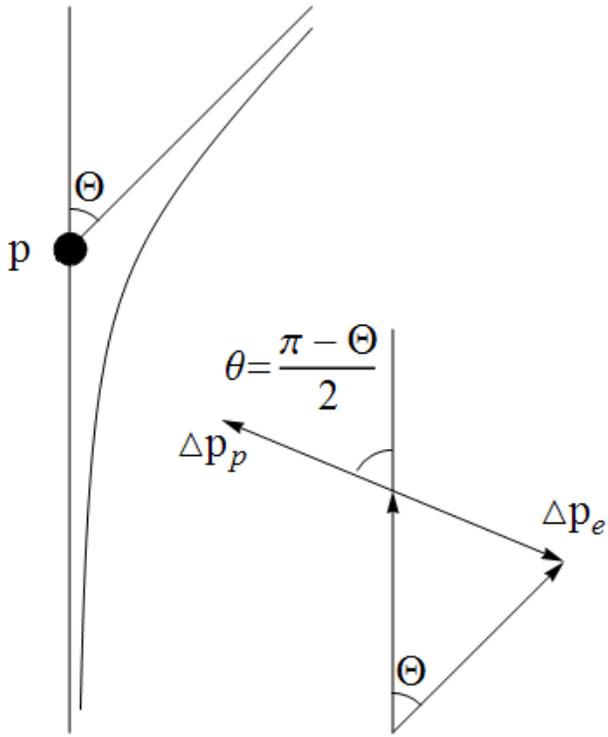}
%\epsscale{.50}
%\plotone{Fig.2.eps}
\caption{The schematic of  collision of a relativistic lepton with a non-relativistic baryon.  After colliding with a baryon at $p$.  The  trajectory of the lepton changes for an  angle, $\Theta $. The exchange of momenta between the lepton and the baryon is shown.\label{fig2}}
\end{figure}

\clearpage

\begin{figure}
\includegraphics[width=0.45\textwidth]{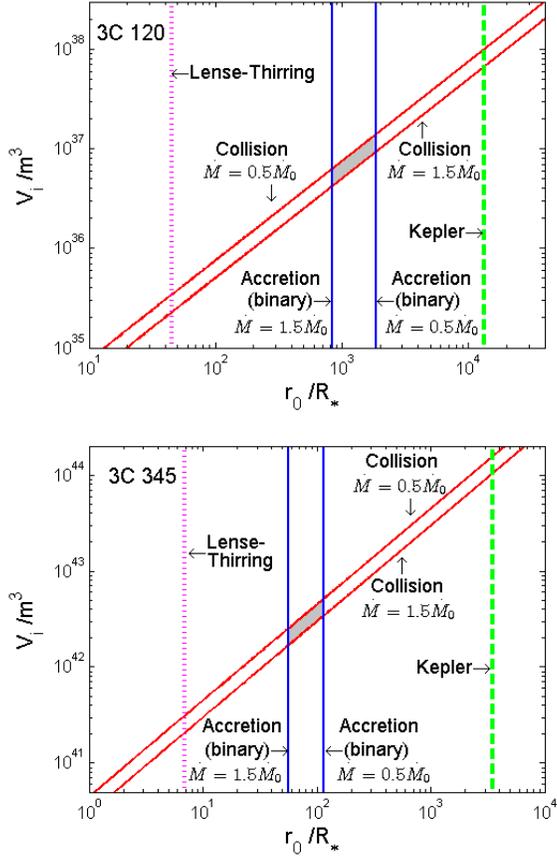}
%\epsscale{.50}
%\plotone{Fig.3.eps}
\caption{Parameters space of ${V_i}$ and ${r_0}$ predicted by the collision model for sources  3C 120 and 3C 345 respectively.
The inclined lines correspond to the ${V_i}$ and ${r_0}$ relation given by Eq.$\ref{Vi}$ with two different values of $\dot{M}$.  The vertical ones in the middle (solid) correspond to ${r_0}$ given by  Eq.$\ref{Ldot1}$ under two different values of $\dot{M}$, which correspond to a collision site given by the tidal torque.
The vertical line at right (dashed) is responsible for  the radius where the Kepler velocity of gases is equal to that inferred from  BLR emission lines.
The vertical line at left (doted) corresponds to the disc radius  when the Lense-Thirring precession yields a precession period of tens of years.\label{fig3}}
\end{figure}

\clearpage

\begin{figure}
\includegraphics[width=0.45\textwidth]{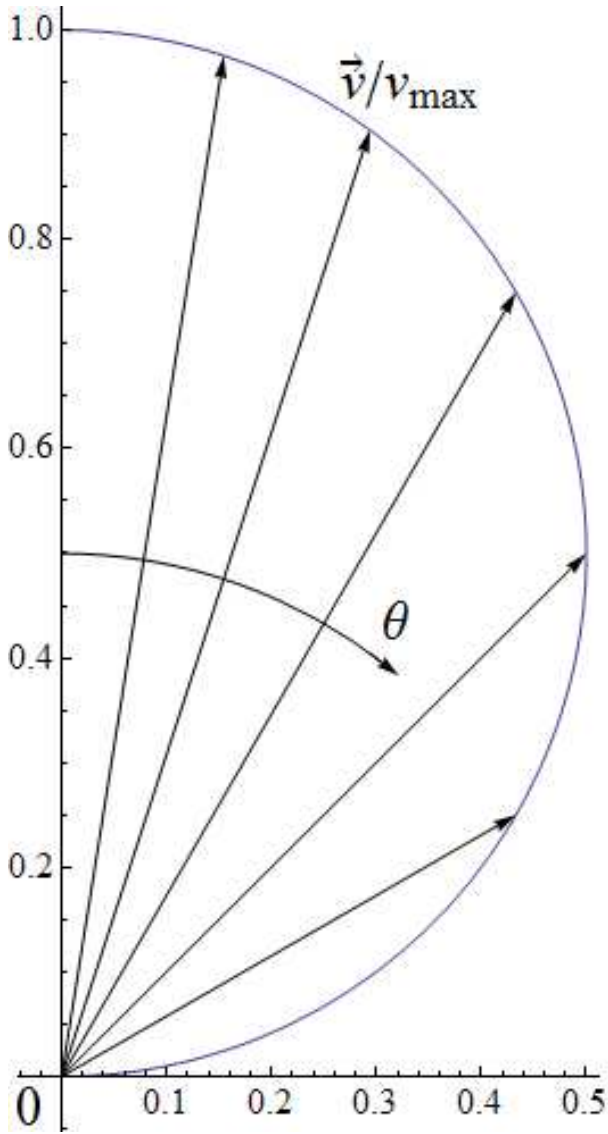}
%\epsscale{.50}
%\plotone{Fig.4.eps}
\caption{The  velocity distribution of scattered baryons. The angle between the  velocity of baryons and the jet axis is $\theta$, the magnitude of the velocity is given by $v = {v_{\max }}\cos \theta $, where ${v_{\max }} = 2\sqrt {2{\gamma _e}} {m_e}c/{m_p}$. Note that the velocity field is axial symmetric about the vertical axis, which predicts a bi-polar configuration of BLR speed.\label{fig4}}
\end{figure}

\clearpage

\clearpage

\begin{table*}%\scriptsize
%\tablecolumns{8} \tablewidth{0pc}
%%\centering
%\begin{minipage}{160mm}
\begin{center}
\caption{\bf Parameters of the Seyfert 1 galaxy 3C 120 and the quasar 3C 345 in the test of jet-disc collision}
\label{Table:Param}
%\resizebox{16cm}{!} {

\begin{tabular}{cccccccccccccccc}
  \hline \hline
  Object & $P(yr)$ & $\mathop M\nolimits_p (\mathop M\nolimits_ \odot  )$ & $\mathop M\nolimits_s (\mathop M\nolimits_ \odot  )$ & $\mathop P\nolimits_{ps} (yr)$ & $\mathop r\nolimits_{ps} (pc)$ & $z$ & $\varphi$ \\
  3C 120 & ${}^a12.3$ & ${}^c3.0 \times \mathop {10}\nolimits^7$  & ${}^c4.0 \times \mathop {10}\nolimits^6$  & ${}^c1.4$ & ${}^c0.002$ & $0.003$ & ${}^a{1.5^ \circ } \pm {0.3^ \circ }$ \\
  3C 345 & ${}^b10.1$ & ${}^c4.4 \times \mathop {10}\nolimits^9$  & ${}^c3.6 \times \mathop {10}\nolimits^9$  & ${}^c5.2$ & ${}^c0.021$ & $0.5928$ & ${}^b{1.3^ \circ } \pm {0.5^ \circ }$ \\
  \hline \hline
\end{tabular}
%}
\end{center}
\begin{flushleft}
{\small NOTE.--$P$ is the precession period measured in the observer's frame, and ${P_{ps}}$ is the orbital period of the binary BH system.  The mass of the primary and secondary BH are  ${M_p}$ and  ${M_s}$ respectively,  ${r_{ps}}$ is the separation between the two BHs, $z$ is the redshift, and $\varphi$ is the half-opening angle of the precession cone given by the model of the following references.\\
%REFERENCES.--a. \cite{CA}; b. \cite{CA1}; c. \cite{CA2}.}
REFERENCES.--a. Caproni \& Abraham 2004b; b. Caproni \& Abraham 2004a; c. Caproni \& Abraham 2004c.}
% or on-off states.  All angels, except for $\rho$, are in radian, and $P_{\rm b}$ is in hour, $c_{1}=S_{1}/L$, $c_{2}=S_{2}/L$,
% and $\phi_{0}$ is the initial phase for $\Theta(t)$. The range in the parenthesis followed each parameter indicates the searching
%  parameters space for the fitting process. The root mean square(rms) of fitting  is given by
% $\sqrt{\sum_{j=1}^N(y^{\prime}_{j}-y_{j})^2/N^{\prime}}$, where $N^{\prime}(=N-10-1)$ is the degree of freedom, $y^{\prime}_{j}$ and $y_{j}$ are the fitted and observed timing residual,respectively. }
 % gives the unweighted $\chi^2$,  obtained in terms of squared deviation between
 %  calculated and observed results per degree of freedom.
\end{flushleft}
%\end{minipage}
\end{table*}

\begin{table*}%\scriptsize
%\tablecolumns{8} \tablewidth{0pc}
%%\centering
%\begin{minipage}{160mm}
%\begin{center}
\caption{\bf The correlation of precession period versus the FWHM of  H$\alpha$ emission lines}
\label{Table:short-period}
%\resizebox{16cm}{!} {
\begin{tabular}{ccccccccccc}
  \hline \hline
  Object & $z$ & $P(yr)$ & $\mathop P\nolimits_{prec} (yr)$ & $FWHM{\kern 1pt} {\kern 1pt} (km\cdot{\rm{ }}{s^{ - 1}})$  \\
  Arp 102B & $0.024$ & ${}^a2.2$ & $2.148$ & ${}^b14400$  \\
  NGC 1097 & $0.004$ & ${}^a5.5$ & $5.478$ & ${}^b7200$  \\
  3C 345 & $0.593$ & ${}^a10.1$ & $6.340$ & ${}^c3600$  \\
  OJ 287 & $0.306$ & ${}^a11.6$ & $8.882$ & ${}^d3710$  \\
  3C 120 & $0.033$ & ${}^a12.3$ & $11.907$ & ${}^c1846$  \\
  3C 273 & $0.158$ & ${}^a16.0$ & $13.817$ & ${}^c3260$  \\
  3C 279 & $0.536$ & ${}^a22.0$ & $14.323$ & ${}^c1400$  \\
  \hline \hline
\end{tabular}
%}
%\end{center}
\begin{flushleft}
%\\
{\small NOTE.--$z$ is the redshift, and $P $ is the precession period measured in the observer's reference frame $P = (1 + z)\mathop P\nolimits_{prec} $.\\
%REFERENCE.--a. \cite{CHA}; b. \cite{WZ}; c. \cite{Roka}; d. \cite{DDT}.}
REFERENCE.--a. Caproni et al. 2004; b. Wang \& Zhang 2003; c. Rokaki et al. 2003; d. Decarli et al. 2011.}
% or on-off states.  All angels, except for $\rho$, are in radian, and $P_{\rm b}$ is in hour, $c_{1}=S_{1}/L$, $c_{2}=S_{2}/L$,
% and $\phi_{0}$ is the initial phase for $\Theta(t)$. The range in the parenthesis followed each parameter indicates the searching
%  parameters space for the fitting process. The root mean square(rms) of fitting  is given by
% $\sqrt{\sum_{j=1}^N(y^{\prime}_{j}-y_{j})^2/N^{\prime}}$, where $N^{\prime}(=N-10-1)$ is the degree of freedom, $y^{\prime}_{j}$ and $y_{j}$ are the fitted and observed timing residual,respectively. }
 % gives the unweighted $\chi^2$,  obtained in terms of squared deviation between
 %  calculated and observed results per degree of freedom.
\end{flushleft}
%\end{minipage}
\end{table*}

\end{document}